\documentclass[showpacs,prl,aps,twocolumn]{revtex4}
\usepackage{graphicx}
\usepackage{amsmath}
\usepackage{bm}
\def\gapx{\lower 2pt \hbox{$\buildrel>\over{\scriptstyle{\sim}}$}}
\def\lapx{\lower 2pt \hbox{$\buildrel<\over{\scriptstyle{\sim}}$}}
\begin{document}
\widetext
\title{Superfluidity of isotopically doped  parahydrogen clusters}

\author{Fabio Mezzacapo and Massimo Boninsegni}

\affiliation{Department of Physics, University of Alberta, Edmonton, Alberta, Canada T6G 2G7}
\date{\today}
\begin{abstract}
It is shown by computer simulations that  superfluid {\it para}-hydrogen clusters of more than 22 molecules can be turned insulating and ``solidlike" by the replacement of  as few as one or two molecules, with ones of the heavier {\it ortho}-deuterium  isotope. A much smaller effect is observed with substitutional {\it ortho}-hydrogen. Substitutional {\it ortho}-deuterium molecules prevalently sit in the inner part of the cluster, whereas {\it ortho}-hydrogen impurities reside primarily in the outer shell, near the surface.
Implications on the superfluidity of pure {\it para}-hydrogen clusters are discussed. 
\end{abstract} 


\maketitle
Clusters of molecular hydrogen are the subject of intense experimental investigation, especially after  the claim of observation of superfluidity (SF) in clusters of  $N$=14-16 {\it p}-H$_2$ molecules surrounding a linear carbonyl sulfide (OCS) impurity \cite{grebenev00}.   Experimental data for pristine or isotopically mixed clusters are not yet available, but novel techniques based on Raman spectroscopy may soon be able to characterize their superfluid properties \cite{toennies04}.

Quantitative information and physical insight in the properties of clusters of {\it para}-hydrogen ({\it p}-H$_2$) molecules, both pristine \cite{sindzingre,guardiola06,roy06,noi06,noi07,buffoni} or doped \cite{kwon02_05, saverio05,paesani,jiang,sebastianelli}, has been provided by microscopic calculations based on Quantum Monte Carlo (QMC) simulations,  making use of accurate intermolecular potentials \cite{SG,silvera,buck84}. In particular, a significant superfluid response has been predicted at $T$ $\le$ 1 K for pristine clusters comprising as many as 27 {\it p}-H$_2$  molecules \cite{noi06,noi07}. 

To an excellent approximation, {\it p}-H$_2$ molecules, as well as those of the heavier {\it ortho}-deuterium isotope ({\it o}-D$_2$) can be regarded as point particles of spin S=0, {\it o}-D$_2$ having twice the mass of {\it p}-H$_2$. 
A third isotope exists, namely {\it ortho}-hydrogen ({\it o}-H$_2$), which has the same mass of {\it p}-H$_2$, but different spin (S=1). 
Pair-wise interactions between molecules of the different isotopes are very nearly the same
\cite{zoppi}; thus, a mixed cluster of  these components provides a simple experimental realization of a mixture of isotopic bosons. A number of fundamental questions can therefore be addressed by studying these systems, e.g., {how does the presence of one component affect the superfluid properties of the other(s), or the structure of the cluster ?}

Classical binary  Lennard-Jones (LJ) clusters have been extensively investigated; for example, it is known that, for equal concentrations of the two species, the one with the smaller depth of the pair potential well $\epsilon_{LJ}$ and/or with higher particle effective radius $\sigma_{LJ}$ will sit prevalently at the cluster surface 
\cite{clarke93,clarke94}. Because in a  mixture of, e.g.,  {\it p}-H$_2$ and {\it o}-D$_2$, $\epsilon_{LJ}$ and $\sigma_{LJ}$ are essentially the same for all pair-wise interactions, the physics of the system  is dominated by quantum effects, at low temperature. 

Path Integral Monte Carlo simulations of such mixed clusters have been carried out down to $T$=2.5 K,  but without including quantum exchanges \cite{chakravarty95}. No theoretical predictions are presently available for the superfluid behavior of this system.  Clearly, the sensitivity of the superfluid response of a finite cluster to its detailed isotopic composition, has a direct relevance to the design and interpretation of ongoing and future experiments.

In this paper,  we present results of a microscopic study of {\it p}-H$_2$ clusters doped with isotopic  impurities; specifically, we investigate the effect of the substitution of few ($N_D \le 4$) {\it p}-H$_2$ with {\it o}-D$_2$ (or, {\it o}-H$_2$) molecules, in ({\it p}-H$_2$)$_N$ clusters that are entirely (or, almost entirely) superfluid. For definiteness, we focus on clusters with $N\ge$ 16,  at  the two temperatures $T$=0.5 and 1 K. 
 
The substitution of  {\it p}-H$_2$ with {\it o}-D$_2$ molecules, generally causes  pristine superfluid {\it p}-H$_2$ clusters  to turn progressively solidlike, their structure increasingly mimicking that of pure {\it o}-D$_2$ clusters with the same numbers of molecules.  For clusters comprising more than 20 molecules, however, the change from liquid to solidlike can occur abruptly. For instance, while the superfluid fraction of the pristine ({\it p}-H$_2$)$_{25}$ cluster at $T$=1 K is approximately 75\%, substitution of a single {\it p}-H$_2$ molecule with an {\it o}-D$_2$ suppresses SF  almost completely. Likewise, two such substitutional {\it o}-D$_2$ impurities suffice to suppress the superfluid response (close to 100\%) of the same cluster at $T$=0.5 K. 

A much smaller reduction of the superfluid response is observed if substitutional impurities are {\it o}-H$_2$ molecules. For example, ({\it p}-H$_2$)$_{24}$--({\it o}-H$_2$)$_1$ is largely liquidlike at $T$=1 K. A crucial observation is that few substitutional {\it o}-D$_2$ molecules sit in the inner part of the cluster, whereas {\it o}-H$_2$ impurities are primarily located near its surface. These findings underscore the prominent role played by long exchanges, especially involving molecules in different shells of the cluster,  in stabilizing a liquidlike structure of large clusters at low $T$,  in turn allowing for SF to occur.

We model our system of interest as a collection of $N_H$  {\it p}-H$_2$  and $N_D=N-N_H$ {\it o}-D$_2$  ({\it o}-H$_2$) molecules, regarded as Bose particles  of spin S=0 (S=1), interacting via the Silvera-Goldman (SG) pair potential \cite{SG}.  
We study this system by QMC simulations, using the continuous-space Worm Algorithm \cite{MBworm,worm2}; technical details of our simulations are the same as in Ref. \cite{noi07}.

 \begin{figure}
\centerline{\includegraphics[scale=0.32, angle=-90]{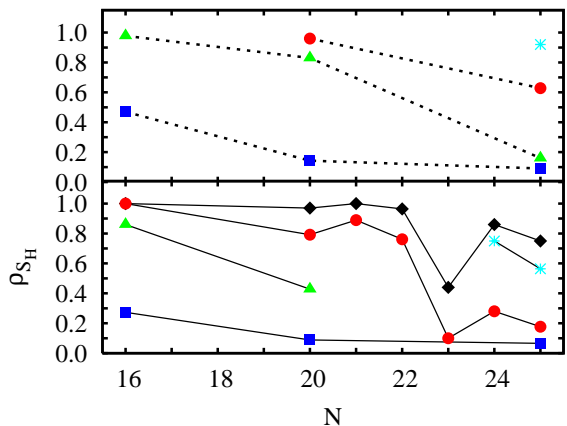}}
\caption{(Color online) Superfluid fraction $\rho_{S_{H}}$  of the {\it para}-hydrogen component in   clusters of $N$  molecules with one (circles), two (triangles) and four (squares) substitutional {\it ortho}-deuterium molecules,  and with one (stars) {\it ortho}-hydrogen molecule. Upper panel shows data at $T$ = 0.5 K, lower at $T$ = 1 K. Diamonds show results for the undoped clusters (entirely superfluid at $T$=0.5 K). Error bars  are comparable to the size of the symbols.  }
\label{ph2rhos}
\end{figure}

Figure \ref{ph2rhos}  shows the superfluid fraction $\rho_{S_{H}}$  of the {\it p}-H$_2$ component, as a function of the cluster size $N$  at $T$ = 1 K (lower panel) and $T$ = 0.5 K (upper panel). Different symbols refer to a different number of substitutional {\it o}-D$_2$  or {\it o}-H$_2$ molecules (see figure caption). 
When a single {\it p}-H$_2$ molecule is replaced by an {\it o}-D$_2$ one, $\rho_{S_{H}}$ is relatively little affected for $N <$ 22, with respect to a pristine {\it p}-H$_2$ cluster, but is depressed substantially in larger clusters, particularly at higher $T$. For example, the substitution of  one {\it o}-D$_2$ in a ({\it p}-H$_2$)$_{25}$ cluster causes $\rho_{S_{H}}$ to drop from 75\% to less than 20\% at $T$=1 K, while at $T$ = 0.5 K an almost complete suppression of the superfluid fraction is observed, in the same cluster, when two {\it p}-H$_2$ molecules are replaced by {\it o}-D$_2$. It should also be noted that pristine {\it p}-H$_2$ clusters with more than 22 molecules, generally display solidlike behavior even when undoped, albeit in different degrees [e.g.,  ({\it p}-H$_2)_{23}$ at $T$=1 K] \cite{noi06,noi07}. 

\begin{figure}
\centerline{\includegraphics[scale=0.32, angle=-90]{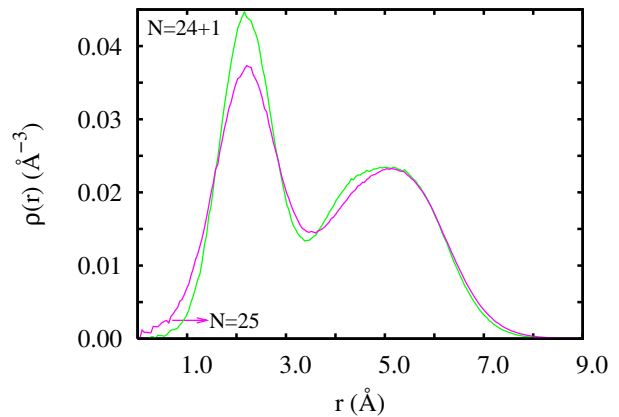}}
\caption{(Color online) Radial density profiles for a pure ({\it p}-H$_2$)$_{25}$ 
(lower peak at short $r$) and a  ({\it p}-H$_2$)$_{24}$--({\it o}-D$_2$)$_{1}$ cluster, at $T$=1 K. Profiles are computed with respect to the geometrical center of the cluster. In the case of the doped cluster, no distinction is made between molecules of different types. Statistical errors are not shown for clarity; they are of the order of 5.0$\times 10^{-4}$ \AA$^{-3}$.}  
\label{radpal}
\end{figure}

Figure \ref{radpal} compares the radial density profile (at $T$=1 K) of the (largely superfluid) pristine ({\it p}-H$_2$)$_{25}$ cluster, with that of the (essentially insulating) cluster with the same number of molecules, but with one {\it p}-H$_2$ replaced by an {\it o}-D$_2$. Profiles are computed with respect to the geometrical center of the cluster \footnote{In the case of a mixed cluster, no distinction is made between molecules of different types.}. The main structural change arising from the substitution of a {\it p}-H$_2$ clearly occurs in the center of the cluster, where molecules are considerably more localized (hence, the higher peak). In the vicinity of the surface of the system, density profiles are, in fact, almost identical. 

If  the dopant is {\it o}-H$_2$, on the other hand, then the structure of the cluster and its superfluid properties are much less sensitive to the substitution (see Fig. \ref{ph2rhos}). 
For example, the corresponding profile for the cluster doped with one {\it o}-H$_2$ molecule is indistinguishable from that of the pristine cluster, on the scale of Fig. \ref{radpal}. 

\begin{figure}
\centerline{\includegraphics[scale=0.34,angle=-90]{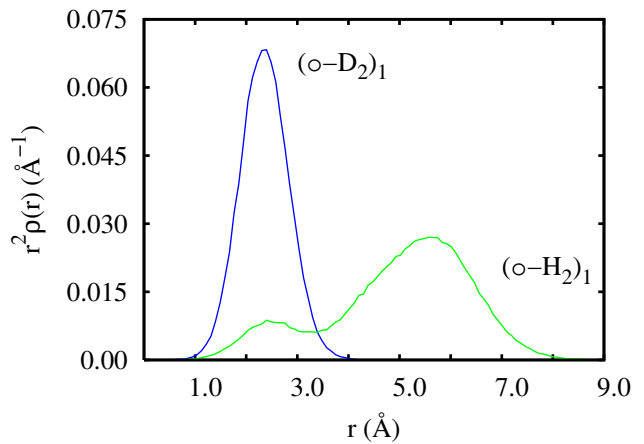}}
\caption{(Color online) Radial density (multiplied by $r^2$)  for the single {\it o}-D$_2$ and {\it o}-H$_2$ impurities in the doped  ({\it p}-H$_2$)$_{24}$--({\it o}-D$_2$)$_{1}$ and ({\it p}-H$_2$)$_{24}$--({\it o}-H$_2$)$_{1}$ clusters, at a temperature of 1 K. Profiles are computed with respect to the geometrical center of the cluster.}  
\label{parort}
\end{figure}

An important observation is that a lone {\it o}-D$_2$ molecule sits in the central part of the cluster; this is something that we observe for all cluster studied, with up to four {\it o}-D$_2$ substitutional impurities, in agreement with previous work \cite{chakravarty95}.  The lighter {\it o}-H$_2$ dopant, conversely, is considerably more delocalized, and indeed is found prevalently in the external part of the cluster, as shown in Fig. \ref{parort}.  

\begin{figure}
\centerline{\includegraphics[scale=0.55]{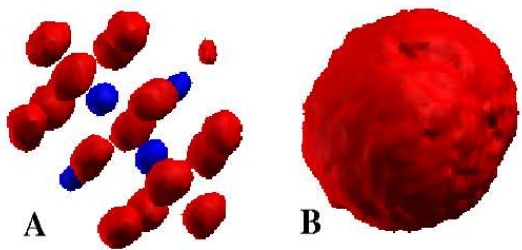}}
\caption{(Color online) Three-dimensional representations of the clusters ({\it p}-H$_2$)$_{16}$--({\it o}-D$_2$)$_{4}$ (A) and of  ({\it p}-H$_2$)$_{20}$ (B). Darker color is used for impurity molecules.}  
\label{ripallocchi}
\end{figure}

The  effect of cluster ``crystallization," induced in relatively large clusters by one or two {\it o}-D$_2$ impurities (and the ensuing depression of  {\it p}-H$_2$ SF)  can also be observed in smaller systems, but  a greater number of substitutions is needed. Figure \ref{ripallocchi} shows the structures of the two clusters  ({\it p}-H$_2$)$_{16}$--({\it o}-D$_2$)$_{4}$ (part A of the figure), and 
({\it p}-H$_2$)$_{20}$ (part B); these pictures were obtained using the procedure   outlined in Ref. \cite{saverio05}. The pristine cluster has a featureless structure, and is entirely superfluid at $T \le$ 1 K. On the other hand, $\rho_{S_{H}}$ is small in the doped cluster, whose solidlike structure is evident, 
with a central axis surrounded by rings of molecules. Two of the four {\it o}-D$_2$ molecules are placed on the axis, the other two on the central ring.

Numerical studies by other authors \cite{saverio05,sebastianelli} had already yielded evidence of  localization  of  {\it p}-H$_2$ molecules around a heavy impurity, rendering small, pristine clusters [e.g.,  ({\it p}-H$_2$)$_{13}$] significantly more rigid and solidlike.
In all previous works, however, impurities were considered such as CO, or HF, not only significantly  heavier, but, more importantly, featuring a stronger (more attractive) interaction with the {\it p}-H$_2$ molecules, than that between the molecules themselves. 

The results presented here, offer insight  in the microscopic mechanism of SF in quantum clusters. In order for SF to occur, clusters (either doped or pristine)  must be essentially liquidlike in character, i.e., molecules must enjoy a high degree of mobility and delocalization.  The most important structural difference between clusters that are insulating or superfluid at $T \le 1$ K, also based on the results for pristine clusters \cite{noi06,noi07}, is that the former feature a rigid, solidlike core, possibly with some loosely bound molecules on the surface; on the other hand, SF is enhanced in clusters whose inner region is floppy, with exchanges taking place between molecules in the  inner and outer shells.

This is consistent with the notion of {\it quantum melting} of large clusters, at low $T$ \cite{noi06,noi07},  originating from permutational exchanges involving {\it all} molecules, including those located in the inner part of the cluster. These exchanges become increasingly important at low temperature, where clusters become increasingly liquidlike, and consequently superfluid.

\begin{figure}
\centerline{\includegraphics[scale=0.33, angle=-90]{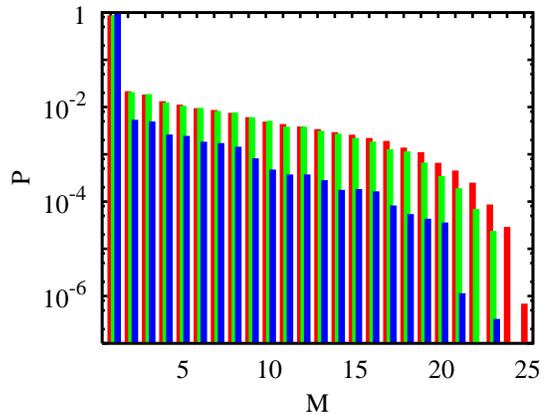}}
\caption{(Color online) Frequency of occurrence of exchange cycles of length $M$ (i.e., involving $1 \le M \le N_H$
{\it p}-H$_2$ molecules) at $T$=1 K. Spikes from left to right refer to a pristine ({\it p}-H$_2$)$_{25}$  cluster and to  the mixtures ({\it p}-H$_2$)$_{24}$--({\it o}-H$_2$)$_{1}$ and ({\it p}-H$_2$)$_{24}$--({\it o}-D$_2$)$_{1}$ . }  
\label{cicli}
\end{figure}

Figure \ref{cicli} displays the frequency with which exchange cycles of
varying length (i.e., involving a different number $1 \le M \le N_H$ of {\it p}-H$_2$) occur (spikes from left to right) in a pristine ({\it p}-H$_2$)$_{25}$ cluster  as well as in the mixtures ({\it p}-H$_2$)$_{24}$--({\it o}-H$_2$)$_{1}$ and ({\it p}-H$_2$)$_{24}$--({\it o}-D$_2$)$_{1}$ at $T$ = 1 K. All exchange cycles are clearly suppressed in the  cluster doped with a single {\it o}-D$_2$, but the reduction is most dramatic for very long cycles.  Conversely, when the cluster is doped with an {\it o}-H$_2$, exchanges (other than the very longest one) take place at almost the same frequency as in the pristine cluster.

Thus, a single substitutional {\it o}-D$_2$, which is located in the center of the cluster owing to its greater mass, has a strong inhibiting effect on  long exchanges of {\it p}-H$_2$ molecules. As a result, the cluster turns solidlike, and SF is altogether depressed. Conversely, a single {\it o}-H$_2$ dopant molecule can effectively get out of the way, thereby allowing for a greater occurrence of long exchanges, including those involving {\it p}-H$_2$ molecules in the inner and the outer shell of the cluster. As a result, the doped ({\it p}-H$_2$)$_{24}$--({\it o}-H$_2$)$_{1}$ cluster remains largely  liquidlike, with a value of $\rho_{S_{H}}$ close to 60 \%.

The suggestion that SF is underlain, or strongly enhanced by the occurrence of exchange cycles involving also molecules located in the central region of the cluster is {not} (as one may naively think) in contradiction with the observation that the {superfluid density} is largest at the surface \cite{buffoni}. For, as discussed above  the primary mechanism by which these exchanges promote SF is {not} by {\it locally} increasing the value of $\rho_{S_H}$, but by stabilizing an overall liquidlike phase of the whole cluster. 

Summarizing, we have studied mixed hydrogen clusters of various sizes, down to $T$=0.5 K. To our knowledge, this is the first study of a mixed isotopic bosonic cluster including all quantum-mechanical effects, namely zero-point motion and permutations of identical particles. Gaining understanding of the effect of substitutional impurities, also affords insight into the microscopic origin of SF in pristine clusters of molecular hydrogen. 

We find that the superfluid fraction $\rho_{S_{H}}$ of  pure {\it p}-H$_2$ clusters is depressed by replacement of few {\it p}-H$_2$ by {\it o}-D$_2$ molecules. The depression is most dramatic for clusters with more than 22 molecules, which display incipient solidlike behavior even when pristine.
The  reduction of the superfluid fraction is considerably smaller in the presence of substitutional {\it o}-H$_2$ impurities. Lighter impurities, which are delocalized throughout the system, have less disruptive an effect on long exchanges than the heavier impurities, which sit in the inner part of the cluster. Our findings are consistent with the notion of melting of large clusters at low temperature as arising from purely quantum-mechanical effects, namely long exchanges of indistinguishable particles.

This work was supported  by the Natural Science and Engineering Research Council of Canada under research grant 121210893, and by the Informatics Circle of Research Excellence (iCORE). 
Simulations were performed on the Mammouth cluster at University of Sherbrooke (Qu\'ebec, Canada).
 One of us (MB) gratefully acknowledges the hospitality of the Theoretical Physics Institute, ETH, Z\"urich.

\end{document}